\documentclass[aps,pre,twocolumn,superscriptaddress,showpacs,showkeys]{revtex4-1}
\usepackage{amssymb}
\usepackage{amsfonts}
\usepackage{rotating}
 \usepackage{amsmath}
\usepackage{graphics}
\usepackage{epsfig}

\usepackage[dvipsnames,usenames]{xcolor}

\begin{document} 
\flushbottom
\title{Committed activists and the reshaping of status-quo social consensus}

\author{Dina Mistry} 
\affiliation{Laboratory for the Modeling of Biological and Socio-technical Systems, Northeastern University, Boston MA 02115 USA} 

\author{Qian Zhang} 
\affiliation{Laboratory for the Modeling of Biological and Socio-technical Systems, Northeastern University, Boston MA 02115 USA}

\author{Nicola Perra} 
\affiliation{Centre for Business Network Analysis, University of Greenwich, Park Row, London SE10 9LS, United Kingdom}
\affiliation{Laboratory for the Modeling of Biological and Socio-technical Systems, Northeastern University, Boston MA 02115 USA}

\author{Andrea Baronchelli} 
\affiliation{Department of Mathematics, City University London, London EC1V 0HB, UK}

\begin{abstract}
The role of committed minorities in shaping public opinion has been recently addressed with the help of multi-agent models. However, previous studies focused on homogeneous populations where zealots stand out only for their stubbornness. Here, we consider the more general case in which individuals are characterized by different propensities to communicate. In particular, we correlate commitment with a higher tendency to push an opinion, acknowledging the fact that individuals with unwavering dedication to a cause are also more active in their attempts to promote their message. We show that these \textit{activists} are not only more efficient in spreading their message but that their efforts require an order of magnitude fewer individuals than a randomly selected committed minority to bring the population over to a new consensus. Finally, we address the role of communities, showing that partisan divisions in the society can make it harder for committed individuals to flip the status-quo social consensus.

\end{abstract}


\maketitle

\section{Introduction}

Social change is often produced by committed groups that challenge the status-quo \cite{harary1959criterion,tilly1978mobilization}. 
Sometimes the transformation is significant, and the new social order is a-posteriori considered an improvement over the old regime. Examples include universal suffrage \cite{przeworski1993political}, the abolition of the transatlantic slave trade, and racial desegregation in the public sphere\cite{chong1991collective}. More often, the change consists of the emergence of a new social consensus on opinions and behaviors that do not seem to be better than the ones they replaced. This is the case we will consider in the present paper. A prominent example concerns shifts in the adoption of social conventions, which by definition are arbitrary behaviors or rules of action \cite{lewis_convention}. This is the case of the constant renewal of current day slang \cite{wentworth1967dictionary}, cultural fads, and fashion \cite{bikhchandani1992theory}. In general, the reshaping of consensus can be extremely fast, and the coexistence between supporters of the old and new status-quo short-lived. 

A natural question concerns the dynamics leading a minority opinion backed by committed supporters to become dominant. Insight into this problem was recently obtained by theoretical work and multi-agent modeling \cite{lu2009naming,xie2011social,marvel2012encouraging,galam_jacobs2007,mobilia_zealots2007, yildiz_2011, mobilia_singlezealot2003, mobilia_voting_2005}. In particular, refs \cite{lu2009naming,xie2011social,marvel2012encouraging} introduced commitment in the context of the Naming Game model of convention formation \cite{Baronchelli_JStatMech_2006,steels1995self}, which has recently been shown to reproduce accurately experimental results on the spontaneous emergence of conventions \cite{Centola_conventions2015}. The model allows agents to hold more than one opinion at the same time, and thus describes `undecided' or `neutral' agents naturally. 
In the model \cite{Baronchelli_JStatMech_2006}, agents interact in pairs, chosen uniformly at random; one of them playing as speaker and the other as hearer, or listener. The speaker randomly selects one of her opinions and transmits it to the hearer. If the hearer holds it in her list, then both speaker and listener retain only that opinion. Otherwise, the listener adds the opinion to her inventory. 
Thus, when the number of opinions is constrained to two, agents can be divided into three groups, namely those who hold opinion $A$, those who hold opinion $B$ and those who hold both opinions, $AB$ \cite{Baronchelli_2007}. A committed minority of individuals that only retain and propagate one opinion can easily flip a majority of individuals initially holding the other opinion provided that its size exceeds a critical value $p_c \simeq 10\%$  \cite{lu2009naming,xie2011social}. Interestingly, this value also holds for the general Naming Game case of $O(N)$ opinions in the system \cite{waagen2015effect} and for a different model on interdependent networks \cite{halu2013connect}.

However, the majority of previous studies did not consider that committed individuals are not only less prone to abandon the opinion they have, but they are usually also more \textit{active} in trying to convince other people \cite{downton1998persistent,young2001activist,botetzagias2010active, borge2013emergence}. Activity-driven networks appear particularly suitable to take these \textit{activists} into account, as they attach to each node a variable, called ``activity'', that describes the propensity of the node to establish new connections at a given time \cite{perra2012activity}. Committed activists are then easily described by assigning a large value of activity to them. It is important to notice that in the classic Naming Game each node is selected uniformly at random, a scenario that corresponds to a homogenous distribution of activity. 

In this paper, we consider the Naming Game model with committed activists on activity driven networks. To this end, we first study a variant of the Naming Game in which only the listeners update their opinion post interaction \cite{Baronchelli_feedback2011}. The listener only model allows us to separate the role of speaker and listener, which is useful in activity driven networks where multiple nodes can be speakers at the same time. 
In the first stage we consider a system characterized by homogenous activity. We show analytically that the threshold of the minimal required committed minority $p_c$ for the listener only variant is around $7\%$ of the population (as estimated in \cite{zhang2011social}), i.e. smaller than the $\simeq 10\%$ obtained when both agents negotiate their position \cite{lu2009naming,xie2011social}. Then, we extend our analysis considering a population of agents described by a heterogeneous distribution of activity. If the activity of the individuals is not correlated to their role (i.e., to be committed or not), the threshold $p_c$ turns out to be the same as found in the previous case where each node is selected uniformly at random.  
Interestingly, we show that a much smaller minority can quickly influence the whole system when committed individuals are more active than the rest of the population. Finally, we consider how a polarized social network, where individuals thinking alike tend to be more connected with each other than to individuals retaining a different opinion, can hinder the effectiveness of a committed minority \cite{adamic2005political, lu2009naming}.

The rest of the paper is structured as follows: We start by introducing the Naming Game with committed minorities on temporal networks in Section \ref{sec: committed_min_NG}. Next we investigate the case of activists as the proponents of the new social convention and the extent of their advantage in Section \ref{sec: activists}. 
In Section \ref{sec: community} we consider collective opinion flipping with committed agents on a real network with community structures built from a network of two weakly connected political blog communities. We are able to show that activists hold a clear advantage over a committed minority chosen at random regardless of network structures present. 
Finally we discuss our results in the Conclusion and 
provide an Appendix for a thorough explanation of the approach taken to arrive at our analytical results. 

\section{Committed minorities in the Naming Game}
\label{sec: committed_min_NG}

The microscopic rules of the Naming Game (NG) model are simple \cite{Baronchelli_JStatMech_2006}. At every time step two agents are selected to interact uniformly at random, one as a speaker and the other as a listener. The speaker randomly selects one of her opinions and shares it with the listener. If the listener has the opinion in her inventory, then both agents retain that opinion and forget all others. If not, the listener adds the opinion to her inventory. This process repeats until all agents agree upon one opinion and only that opinion. Agents begin the game with no prior knowledge of any opinions and new opinions are created by speakers who have none to share.

Here, we will focus on the case where only $2$ opinions are possible \cite{Baronchelli_2007}, also known as the binary NG\cite{Baronchelli_feedback2011,xie2011social}. Thus agents belong to one of three possible groups; those with opinion $A$, $m_A$, those with opinion $B$, $n_B$, or those with both $A$ and $B$, $n_{AB}$, such that $m_A + n_B + n_{AB} = 1$. With $A$ as the opinion of the committed minority, the group holding opinion $A$ can be further split into two; those who can be influenced and persuaded to adopt opinion $B$, $n_A$, and zealots committed 
to opinion $A$, $p$, so that $m_A = n_A + p$. 

In the NG, and also in previous studies with committed individuals, both speakers and listeners negotiate their opinion state for each interaction. In this paper, we will focus on the listener only or hearer only NG 
a variant where only the listener updates their opinion after an interaction, which  yields the same scaling of convergence time with population size $N$ as observed in the usual NG \cite{Baronchelli_feedback2011}.

With our interests lying in the study of collective opinion flipping, we consider the case in which only the committed minority know of the new convention ($A$) at the beginning, while everyone else agrees on the old convention ($B$). Previous studies of the NG with committed minorities considered this scenario resulting in a critical size of the committed minority $p_c \simeq 10\%$ \cite{lu2009naming,xie2011social}. In the hearer only NG variant (hereafter referred to as the NG) we find the critical size becomes smaller with $p_c \simeq 7\%$. This result is derived from a fixed point stability analysis of the mean field rate equations for the densities of agents in the separate states A and B, as detailed below.

\subsection*{Committed minority threshold for the Hearer-Only Naming Game} 
With the density of committed agents fixed to $p$ and all agents having $a$ $priori$ knowledge of at least one opinion, the fraction of the population knowing $A$ and $B$ is given by $n_{AB} = 1 - n_A - n_B - p$. 
The mean field rate equations of the states are then easily obtained by considering the possible interactions listed in Table \ref{tab: outcomes_table}. They read:

\begin{align}
\label{eq: dotnA}
\dot{n_A} &= -n_An_B + \frac{1}{2}n_An_{AB} + \frac{1}{2}n_{AB}^2 + pn_{AB}\\
\label{eq: dotnB}
\dot{n_B} &= -n_An_B + \frac{1}{2}n_Bn_{AB} + \frac{1}{2}n_{AB}^2 - pn_B
\end{align}
\noindent

\noindent 
The terms on the right of (\ref{eq: dotnA}) describe the number of agents leaving and joining the state $n_A$ according to different interaction pairs. In particular, the first considers  individuals who leave $n_A$ after hearing from a speaker with only $B$ in their memory. The second is a combined term describing those who join $n_A$ from $n_{AB}$ after hearing from a speaker with only $A$ in their memory and those who leave $n_A$ after hearing $B$ from a speaker with $A$ and $B$, where both have equal probability to be transmitted. The terms on the right of (\ref{eq: dotnB}) are similar but describing the number of agents leaving and joining state $n_B$.

\begin{table}
\begin{tabular*}{\columnwidth}{@{\extracolsep{\stretch{1}}}*{3}{c}@{}}
  \toprule
  Speaker & Listener & Listener \\
  & before interaction & after interaction \\
  \hline
  $A, A_c$ & $A, AB$ & $A$ \\
   & $B$ & $AB$ \\
   & $A_c$ & $A_c$ \\
  \hline
  $B$ & $A$ & $AB$ \\
   & $B, AB$ & $B$ \\
   & $A_c$ & $A_c$ \\
  \hline
  $AB \xrightarrow{A}$ & $A, AB$ & $A$ \\
  & $B$ & $AB$ \\
  & $A_c$ & $A_c$ \\
  \hline
  $AB \xrightarrow{B}$ & $A$ & $AB$ \\
  & $B, AB$ & $B$ \\
  & $A_c$ & $A_c$ \\                             
\end{tabular*}
\caption{Interaction outcomes for different speaker-listener pairs in the $2$ state (hearer only) NG. For speakers in the state AB there are two possible opinions to share; the opinion shared is indicated above the arrow. $A_c$ refers to agents committed to opinion $A$.}
\label{tab: outcomes_table}
\end{table}

Following the analysis of Xie, et al \cite{xie2011social}, we determine the conditions of existence for the fixed points of the above mean field rate equations. Simplifying our notation by replacing $n_A$ with $x$ and $n_B$ with $y$, we get the following:

\begin{align}
\label{eq: dotx}
\dot{x} = &-xy + \frac{1}{2}x(1-x-y-p)  + \frac{1}{2}(1-x-y-p)^2 
\\&+ p(1-x-y-p)
\nonumber
\\
\label{eq: doty}
\dot{y} = &-xy + \frac{1}{2}y(1-x-y-p) + \frac{1}{2}(1-x-y-p)^2 -py \nonumber
\\& 
\end{align}
\noindent The fixed points of this system correspond to (x,y) points which satisfy $\dot{x} = \dot{y} = 0$, giving us:

\begin{align}
\label{eq: x}
x &= \frac{(1-y)^2 - p^2}{1+y+p}\\
\label{eq: y}
y &= \frac{(1-x-p)^2}{1+x+p}
\end{align}
\\

\noindent With (\ref{eq: dotx}) substituted into (\ref{eq: doty}) we get:
\begin{equation}
\label{eq: yexpression}
2y(1+y+p)\big( 3y^2 + 4(p-1)y + (p+1)^2\big) = 0
\end{equation} 

\noindent The fixed point values of $y$, $y*$, are given by solutions of this expression. 

Neglecting trivial solutions, the third factor in (\ref{eq: yexpression}) gives one or two additional fixed points depending on the parameter $p$:

\begin{equation}
\label{eq: solny}
y* = \frac{-2(p-1)\pm\sqrt{p^2-14p+1}}{3}
\end{equation}

\begin{figure}
\includegraphics[width=\columnwidth] {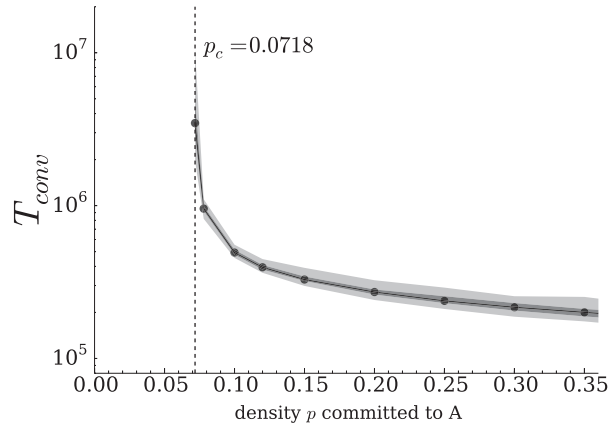}
\caption{
\footnotesize{Numerical results for convergence time $T_{conv}$ in the binary NG with a randomly selected committed minority for a population of $N = 10^4$ agents. Only two opinions are possible: $A$ or $B$, with a state of consensus reached when everyone knows only opinion $A$. Committed agents begin with knowledge of $A$ and introduce this new opinion to the rest of the population who initially know opinion $B$. The median of $100$ simulations show that convergence time diverges at the analytically derived critical threshold of $p_c \approx 0.0718$. The $95\%$ and $50\%$ confidence intervals are shown shaded in dark and light grey.
}}
\label{fig: NG_pclimit}
\end{figure}

\noindent The only physical solution gives 
$p_c = 7-4\sqrt{3} \approx 0.0718$, confirming the previous numerical estimate $p_c  = 0.08 \pm 0.01$ of \cite{zhang2011social}.  
At $p_c$ the additional fixed points afforded by (\ref{eq: solny}) collapse to a single point to give the fixed state ($n_A$,$n_B$) = $(0.0829,0.6188)$, a saddle point which splits off into two fixed points for $0\leq p \leq p_c$. 
The existence of fixed points where $y* \neq 0$ indicates the possibility of stable states for the population in which opinion $B$ persists and convergence is never met. 
Then $p_c$ also represents a critical threshold around which convergence time $T_{conv}$ 
diverges (see Figure \ref{fig: NG_pclimit}). 

Figure \ref{fig: NG_pclimit} shows that as $p \rightarrow p_c$, average convergence time begins to diverge in simulations of the NG for a population of $10^4$ agents. At minimum then $\sim7\%$ of the population must be committed to opinion $A$ in order to strike an effective mutiny against the prevailing opinion $B$. 

\subsection*{Committed minorities in heterogeneous populations}  
Thus far we have considered the NG in homogeneous populations, where all individuals have the same probability of speaking. However, in real life social networks are comprised of individuals heterogeneous in their tendency to connect and communicate. To take this into account, we consider activity-driven networks\cite{perra2012activity,karsai2014memory,tomasello2014role}, in which nodes are assigned an activity rate \textit{a} which describes their propensity to communicate. At the start of each time step \textit{t} nodes activate with a probability $a\Delta t$ and connect to \textit{m} random neighbours, where $\Delta t$ is the duration of the step. At the next time step $t+\Delta t$ all links are broken and the network is built anew. For our model we only consider the case where  \textit{active} nodes connect to \textit{inactive} nodes. Activity rates are values in the interval $(\epsilon,1)$, where $\epsilon << 1$, and are given according to a probability distribution function $P(a)$. For a wide set of large datasets describing social interactions the function $P(a)$ is shown to be a power law function of $a$ such that $P(a) \propto a^{-\gamma}$ \cite{perra2012activity,karsai2014memory,ribeiro2013quantifying}. This results in a heavy-tailed activity distribution with a lower activity cut-off $\epsilon$. We will use with this functional form for $P(a)$ to produce populations with wide varying rates of activity from which we will select our \textit{activists}, where $\gamma = 2.5$, $\epsilon = 10^{-3}$, and each active agent connects to $m = 1$ inactive neighbors per time step $\Delta t = 1$. 

We now consider the NG on these time varying networks where speakers are active nodes and listeners are their \textit{m} inactive neighbours. 
In this setting, listeners update their opinion at the end of the time step. 
After all speakers have communicated each listener will retain the mode of their opinions. However, nodes have a probability $\propto N^{-2}$ of being a listener for $s$ speakers during a given time step, so that in large populations listeners typically receive only one opinion in a given time step.

To determine the critical threshold $p_c$ it is convenient to write down equations for each activity rate or class $a$. Thus, $n^{a}_y$ represents the fraction of population whose activity is $a$ which holds opinion $y$, and $p^a$ is the fraction of committed individuals with activity $a$.  The rate equations for the density of states $A$ and $B$ at each class $a$ are:
\begin{align}
\label{eq: dotnA_act}
\dot{n^a_A} = &-n^a_A\sum_{\substack{a'}} n^{a'}_Ba' + 
n^a_{AB}\sum_{\substack{a'}} n^{a'}_Aa'
-\frac{1}{2}n^a_A\sum_{\substack{a'}} n^{a'}_{AB}a' \nonumber 
\\&
+ \frac{1}{2}n^a_{AB}\sum_{\substack{a'}} n^{a'}_{AB}a' + n^a_{AB}\sum_{\substack{a'}} p^{a'}a' & 
\\
\label{eq: dotnB_act}
\dot{n^a_B} = &-n^a_B\sum_{\substack{a'}} n^{a'}_Aa' + n^a_{AB}\sum_{\substack{a'}}n^{a'}_Ba'
-\frac{1}{2}n^a_B \sum_{\substack{a'}} n^{a'}_{AB}a' \nonumber
\\&
+ \frac{1}{2}n^a_{AB}\sum_{\substack{a'}} n^{a'}_{AB}a' -n^a_B\sum_{\substack{a'}} p^{a'}a' & 
\end{align}
\noindent 
where each summation is an average over all activity classes for speakers in different states.  For each term in the above the left side describes the state of the listener in class $a$, the right side is the average speaker communicating with them who can cause them to change their position, while the factor in front is the probability for this change. We can simplify these equations with the use of the following notation:
$
n_A = x, 
$
$
n_B = y, 
$ 
$n_{AB} = z = 1 - x - y - p, $ 
$\widetilde X = \sum_{\substack{a}} n^a_Aa,  $
$\widetilde Y = \sum_{\substack{a}} n^a_Ba, $
$\widetilde Z = \sum_{\substack{a}} n^a_{AB}a. $

Summing 
(\ref{eq: dotnA_act}) and (\ref{eq: dotnB_act}) over all classes of $a$ then gives us the full rate equations:

\begin{align}
\label{eq: xdot}
\dot{x} &=  -x\widetilde Y + z\widetilde X -\frac{1}{2}x\widetilde Z + \frac{1}{2}z\widetilde Z + zp\bigl\langle a \bigr\rangle
\\
\label{eq: ydot}
\dot{y} &= -y\widetilde X + z\widetilde Y -\frac{1}{2}y\widetilde Z + \frac{1}{2}z\widetilde Z -yp\bigl\langle a \bigr\rangle
\end{align}

\noindent where $\sum_{\substack{a}} p^aa = p\langle a \rangle$. This comes from committed agents being randomly distributed among activity classes resulting in $p^a$ being proportional to the probability of agent having activity class $a$, thus $p^a = pP(a)$ so that when summed over all classes the total density of committed agents is $p$. 

With some consideration on the definition for z (see Appendix), we find that $\widetilde Z =  \langle a\rangle - \widetilde X - \widetilde Y - p\langle a\rangle$. 
We note the existence of solutions for $x$ and $y$ relies only on the existence of valid solutions for $\widetilde X$ and $\widetilde Y$.
Three steady state solutions are found for $\widetilde X$ and $\widetilde Y$ through a fixed point stability analysis of their rate equations (Appendix: Activity-driven networks).
Adopting the same line of reasoning seen for homogeneously mixing populations, 
we find $p_c = 7 - 4\sqrt{3}$.  

\begin{figure}
\includegraphics[width=\columnwidth] {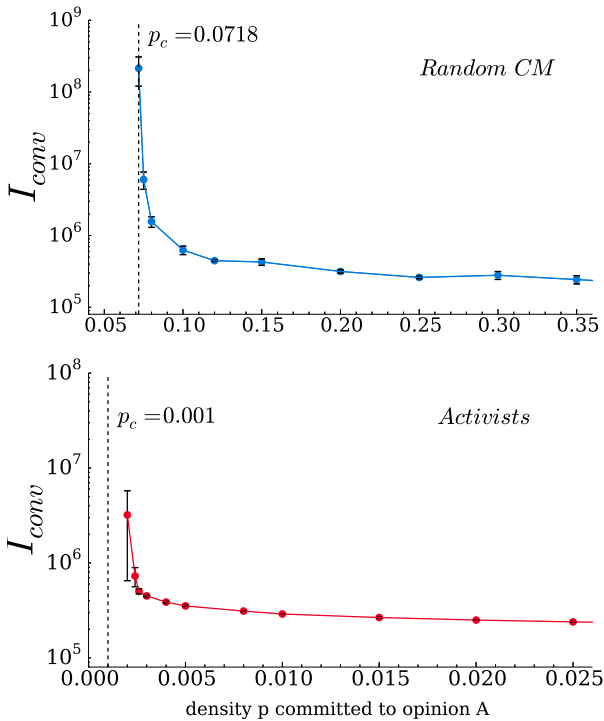}
\caption{\footnotesize{(Color online) Average number of interactions required to reach convergence $I_{conv}$ vs $p$ in the NG in activity driven networks for a population of $N = 10^4$ agents for 100 simulations (Top). 
$I_{conv}$ diverges at the same $p_c$ for random committed minorities as in systems with homogenous activity distributions (Figure 1).
(Bottom) $Activists$ (commitment correlated with high probability to speak) on the other hand show considerable advantage in needing fewer interactions to reach consensus and a numerical threshold of $p_c \approx 2 \times 10^{-3}$, agreeing well with the analytically derived $p_c \approx 1 \times 10^{-3}$. } 
}
\label{fig: Interactions10000}
\end{figure}

This is the same critical threshold for $p_c$ obtained for the case of a homogeneous activity distribution, and numerical simulations support this result, once time is rescaled so as to take into account multiple speakers at each time step rather than a single speaker 
($I_{conv}$ now instead of $T_{conv}$ for the number of interactions needed to reach convergence; 
see Figure \ref{fig: Interactions10000}).  
Thus, the introduction of varying rates of communication through activity driven networks has no effect on $p_c$ when our committed minority is selected at random while $I_{conv}$ increases by less than two orders of magnitude near $p_c$. 

\begin{figure}
\includegraphics[width=\columnwidth]{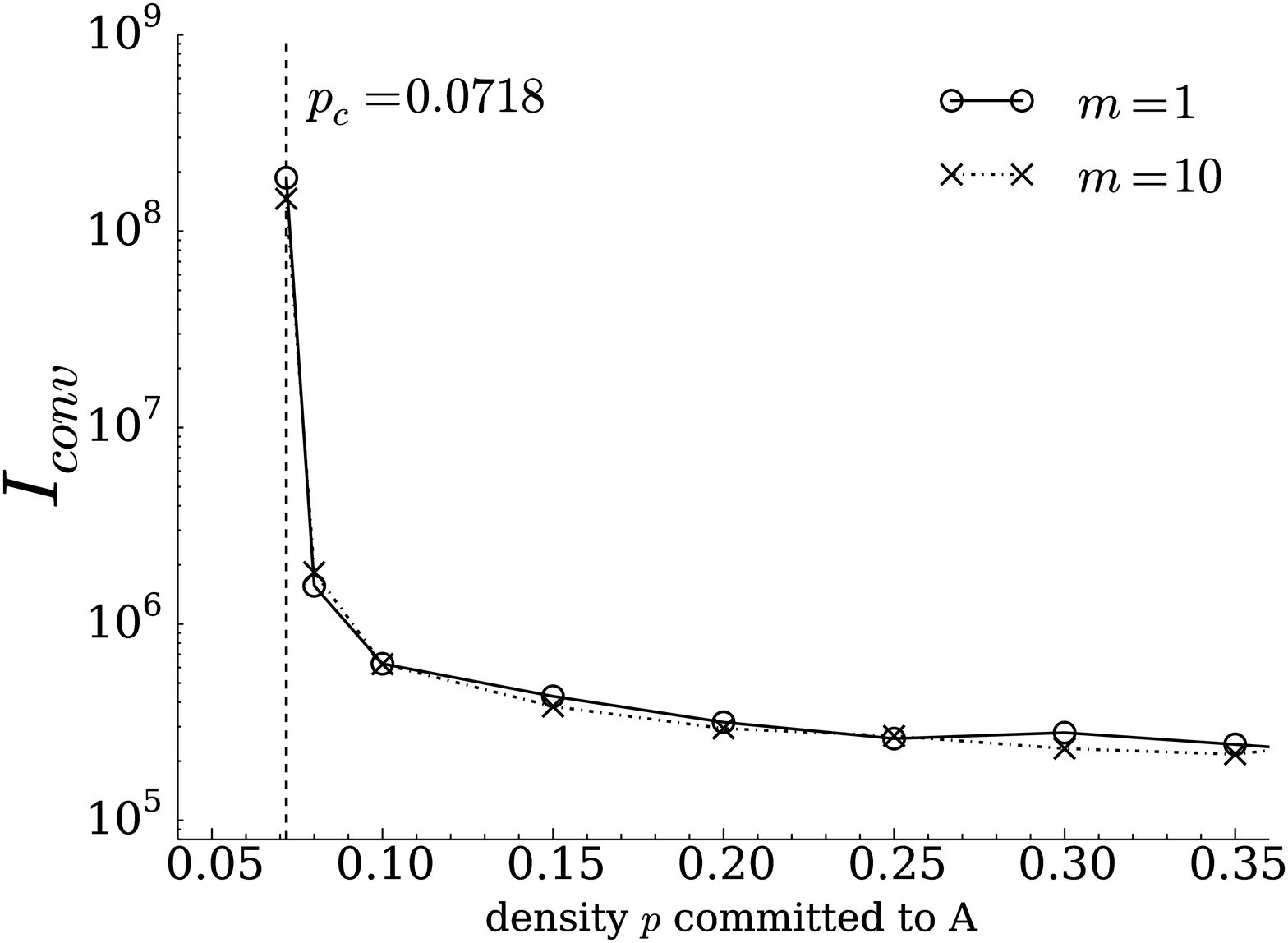}
\caption{\footnotesize
{Comparison of average number of interactions required to reach convergence $I_{conv}$ vs $p$ in the NG in activity driven networks with $N = 10^4$ agents for $m = 1$ and $10$ listeners per speaker per time step. $I_{conv}$ and the critical threshold $p_c$ remains the same despite a tenfold increase in communication per step. Error bars are not visible at the scale of the plot.}
}
\label{fig: Interactions10000_mcurve}
\end{figure}

From Figure \ref{fig: Interactions10000_mcurve} we see that increasing $m$ - the number of neighbors speakers communicate with at each step - by an order of magnitude also has no effect on $p_c$ nor $I_{conv}$ with committed agents selected at random.

\section{Activists}
\label{sec: activists}

Individuals committed to a cause are not only less likely to leave their position but are also far more active in recruiting others for their cause \cite{downton1998persistent,young2001activist,botetzagias2010active}. 
To describe the presence of activists, we select the committed minority to be the most active agents in the population, thus correlating the agent's activity rate $a$ with commitment. These activists will communicate the most often in our population and have many more opportunities to share their message.
Concretely, activists are now chosen to be the $Np$ agents with the highest activity rates $a$. This results in a different definition of a committed agent of activity class $a$. Only a range of high activity classes are now selected according to $p^a\propto P(a)$, while $p^a = 0$ holds for the rest of the population since by definition no activist comes from a low activity class. Hence,

\begin{align}
\label{eq: committed_pa}
p^a = \begin{cases}
 P(a) &\mbox a_c \leq a \leq 1 \\
 0 & \mbox a < a_c
\end{cases}
\end{align}

\noindent where $a_c$ is the lower limit of activity classes for activists defined from the integral:

\begin{equation}
\label{eq: committed_p}
p = \int^1_{a_c} P(a)da,
\end{equation}

\noindent  as the integral of all activists of the higher activity classes must equal the total density of activists $p$ in the population. The lower limit of activity classes for activists is: 

\begin{equation}
\label{eq: lowerlimit_a}
a_c = \left( 1- p(1-\epsilon^{-\gamma+1}) \right)^\frac{1}{-\gamma+1}
\end{equation}

\noindent
The rate equations (\ref{eq: xdot}) and (\ref{eq: ydot}) need to be modified accordingly, resulting in 

\begin{equation}
\label{committed_pa_sum}
\sum_{\substack{a}} p^aa = \int^1_{a_c} P(a)ada = \left(\frac{-\gamma+1}{-\gamma+2}\right)\left(\frac{1 - a_c^{-\gamma+2}}{1 - \epsilon^{-\gamma+1}}\right)
\end{equation}

\noindent We will refer to this term as $\langle a_c\rangle$; it is the average activity of activists. 
Using this $\widetilde Z$ becomes:

\begin{equation} 
\label{Z_act}
\widetilde Z =  \langle a \rangle - \widetilde X - \widetilde Y - \langle a_c \rangle
\end{equation}
 
\noindent and the rate equations for $\widetilde X$ and $\widetilde Y$ become:
 
\begin{align}
\label{eq: X_act}
\dot{\widetilde X} &= -\widetilde X\widetilde Y + \frac{1}{2}\widetilde X\widetilde Z +\frac{1}{2}\widetilde Z^2 + \widetilde Z\bigl\langle a_c\bigr\rangle
\\
\label{eq: Y_act}
\dot{\widetilde Y} &= -\widetilde X\widetilde Y + \frac{1}{2}\widetilde Y\widetilde Z +\frac{1}{2}\widetilde Z^2 -\widetilde Y\bigl\langle a_c\bigr\rangle
\end{align}

\noindent Real solutions for $\widetilde X$ and $\widetilde Y$ yield the value $p_c \approx$ $1.0 \times 10^{-3}$  
for the critical threshold. Below $p_c \approx  1.0 \times 10^{-3}$ then we can expect to find stable states for the population where opinion $B$ never dies out and convergence to $A$ is never reached. Figure \ref{fig: Interactions10000} shows that numerical simulations for the case of $activists$ agree well this theoretical prediction. 
Convergence occurs for a significantly lower range of $p$ compared to a random committed minority where $p_c \approx 0.0718$. The number of interactions needed to reach convergence with activists begins to diverge
around $p \approx 2 \times 10^{-3}$.

\section{Community Structure}
\label{sec: community}

In the real world, the effectiveness of activism might be hindered by the tendency of individuals to communicate preferentially with like-minded peers \cite{mcpherson2001birds,conover2011political}. To investigate this point, we consider the simple case of two weakly connected communities by examining the network of online political blogs 
analysed in Adamic and Glance's work on the polarized political blogosphere two months before the $2004$ U.S. Presidential Election \cite{adamic2005political}, which shows two distinct communities: Democratic and Republican leaning blogs. 

The original dataset constructs a directed network so we will focus on the greatest weakly connected component which contains $1222$ blogs and $19089$ links out of the total $1490$ blogs and $19090$ links and use an undirected representation of this network. 
The remaining network is split nearly into two equal sized communities: $586$
liberal blogs and $636$ conservative blogs. The two community structure of this network then provides a natural division for knowledge of the two opinions in the binary NG with one community being the source for the new social convention. 
For this network then we will focus on two scenarios of the NG:
\begin{enumerate}
\item In the first scenario agents begin with knowledge of opinion $A$ or $B$ dependent on the community they are a part of, with a select group from the first community committed to $A$;
\item In the second scenario everyone knows opinion $B$ except for the group committed to $A$ in the first community. 
\end{enumerate}

The activity is defined as the propensity of each node to engage in social interactions. For each node $i$, it is proportional to the ratio between the number of interactions involving the node and the total number of interactions in the system. Formally, the activity is then $a_i=\eta\frac{s_i}{\sum_{j}s_j}$ where $s_i$ is the strength of $i$, i.e. total number of interaction of $i$, and $\eta$ is a rescaling factor. In the dataset we consider here, we have information just about the number of different peers, the degree $k_i$, in contact with $i$. For this reason, we generate the activity rates from the real data considering the normalized degree for individual blogs or agents in the network  $a_i=\eta \frac{k_i}{\sum_{\substack{i}}k_i}$. At each time step active agents connect to an inactive neighbor chosen among  those for which they have an existing link to in the original dataset. 
This preserves the community structure, allows for an activity-driven creation of links in the network and produces a ranking of agents by which we can select $activists$. In contrast to the activity driven networks previously considered, where the distribution of activity rates were given by a power law with exponent $\gamma = 2.5$, this political blogosphere network exhibits a power law distribution of normalized degrees with $\gamma \approx 1.465 \pm 0.014$. We use $\eta = 50$ to raise the activities so that the number of active agents per time step in this network is 
at least 1.

\begin{figure}
\includegraphics[width=\columnwidth]{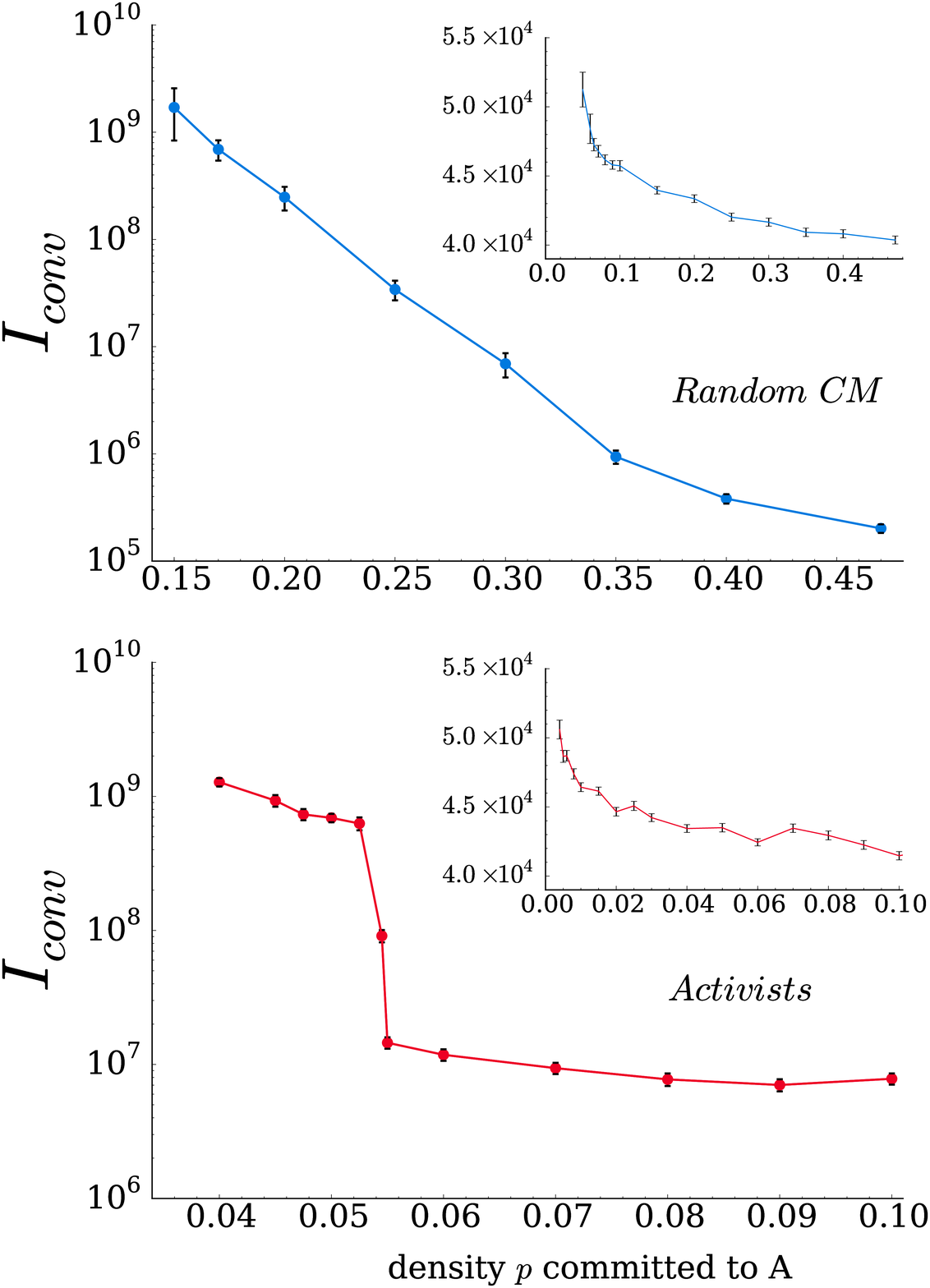}
\caption{\footnotesize{(Color online) $I_{conv}$ in the NG on a real network of $N = 1222$ agents containing two weakly connected communities vs a random rewiring of the network (inset). Agents begin with knowledge of opinion $A$ or $B$ dependent upon the community they are originally a part of with a select group in community $A$ remaining committed to $A$.  
$I_{conv}$ is the number of interactions at which $95\%$ or more of the network agrees upon opinion $A$.
In both the original community based network and the rewired network, shown in the insets, $activists$ (bottom) show considerable advantage over randomly selected committed minorities (top) in needing fewer committed agents.}}
\label{fig: community_interactions} 
\end{figure}

Figure \ref{fig: community_interactions} shows the results of the NG played out in this community structured network in comparison to a random rewiring of the network (shown in the inset)
in the first scenario detailed above. Agents begin with knowledge of opinion $A$ or $B$ depending on the community they are a part of and we select a group within community $A$ to be committed.
$I_{conv}$ is now the number of interactions required to get $95\%$ or more of the network to hold only opinion $A$. We lower the conditions for convergence from $100\%$ of the network holding only opinion $A$ to $95\%$ to guarantee a faster yet accurate estimation of the threshold.

Rewired networks are obtained by reshuffling the end points of the links, which destroys the community structure without altering the degree, and therefore activity, of each blog. The comparison of the results obtained in the real network against those obtained in the rewired network is crucial to highlight and isolate the effects of the community structure.
Numerical results, shown in Figure \ref{fig: community_interactions}, indicate that once again $activists$ demonstrate a considerable advantage over a randomly selected committed minority in both the original two community network ($p_{ca} = 0.04$ vs $p_{cr} = 0.15$, thresholds for activists and a randomly selected committed minority, respectively) and the rewired network shown in the inset ($p_{ca} = 4.0 \times 10^{-3}$ vs $p_{cr} = 0.05$; 5 blogs vs 62 blogs)
with $activists$ needing a significantly smaller minimal committed group to convince the rest of the population to their position. 

\begin{figure}
\includegraphics[width=\columnwidth]{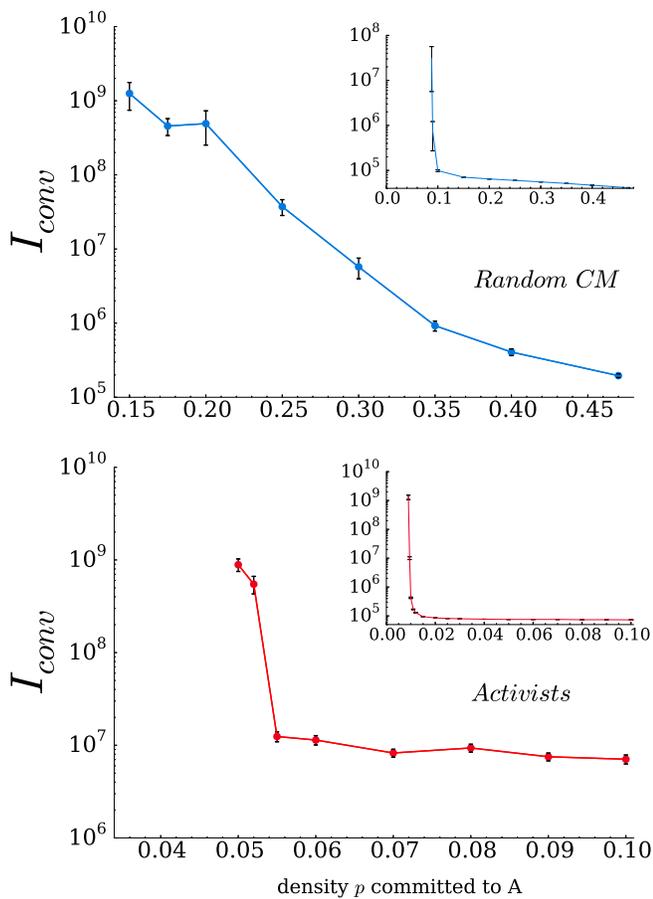}
\caption{\footnotesize{(Color online) $I_{conv}$ in the NG on a real network of $N = 1222$ agents containing two weakly connected communities vs a random rewiring of the network (inset). All agents begin with knowledge of opinion $B$, regardless of their original community membership except for a select group in community $A$ who remain committed to $A$. $I_{conv}$ is the number of interactions at which $95\%$ or more of the network agrees upon opinion $A$.
In both the original community based network and the rewired network, shown in the insets, $activists$ (bottom) show considerable advantage over randomly selected committed minorities (top) in needing fewer committed agents.}}
\label{fig: community_interactions_withB}
\end{figure}

Figure \ref{fig: community_interactions_withB} shows the results of the NG in the second scenario. All agents begin with knowledge of opinion $B$ except for a small segment of the population in one community who introduce opinion $A$ and remain committed to their cause. This scenario is similar to the initial configuration we previously considered for activists in activity-driven networks. Again 
$I_{conv}$ is the number of interactions required to get $95\%$ or more of the network to hold only opinion $A$. Here
activists show the same advantage over a random committed minority, needing a smaller minimal committed group in both the original network ($p_{ca} = 0.05$ vs $p_{cr} = 0.15$) and the rewired network ($p_{ca} = 9 \times 10^{-3}$ vs $p_{cr} = 0.088$).

Finally, Figures \ref{fig: community_interactions} and \ref{fig: community_interactions_withB} also show that the original network (main panels) requires a larger minimal committed group 
to persuade the rest of the population to their position than the rewired networks (insets). Thus, community structures present in the network 
inhibit the spreading of opinion $A$ and hinder the effectiveness of committed agents. 
Even activists in the community structured network need a larger minimum a size of the network (at least $\sim 4\%$) to spread their message, an order of magnitude higher than required in the rewired network for the first scenario.  

\section{Conclusion}
\label{sec: conclusion}

We investigated the role of committed individuals in the Naming Game model on activity-driven networks. 
First, we considered a variant of the Naming Game, in which only the listeners update their opinion, considering a homogenous distribution of activity. Interestingly, we found the critical threshold $p_c$, defining the minimal fraction of committed individuals needed for a fast opinion flipping of the majority, to be smaller than the same value obtained when both agents negotiate their positions. Then we considered the presence of heterogeneous activity patterns in the propensity to communicate. Surprisingly, we found that this characteristic, observed in many networks, does not alter the critical threshold $p_c$. Furthermore, we considered the effects of \textit{activists} by correlating activity, or propensity to speak, with commitment. We found that \textit{activists} can reduce their numbers by two orders of magnitude 
compared to random committed agents and still persuade their peers to adopt their opinion or social convention. Finally, we considered the presence of communities in a real social network. We showed that communities inhibit the ability of committed agents to effectively spread their message and subsequently require larger committed groups to convince the rest of the population to join their position.

Taken together our results indicate that strategically selecting individuals for a given cause or social convention can greatly reduce the cost of associated campaigns in terms of the sheer number of individuals needed for its success. It is worth noticing that an approach similar to the one presented here was proposed in \cite{borge2013emergence} for the rumor spreading model \cite{goffman1964generalization}, with the difference that the authors considered a static network mimicking the fact that agents would remember their connections and explore only a fixed subset of the network.  Recent results on the effects of social memory and the heterogeneity of social ties on spreading phenomena suggests that 
future work may benefit from including more realistic partner selection mechanisms also in temporal networks to better reflect what is observed in real social networks~\cite{karsai2014memory,Karsai2010,miritello2011dynamical,onnela2007structure}. Other interesting points left for future exploration are the influence of fatigue on demobilization and disengagement of activists \cite{beinin2013social, passy2000socialization} (here modeled as endlessly committed), and the role of broadcasting agents able to reach a large part of the population (i.e., mass media) in the spirit of previous studies that addressed this point for Axelrod's model of the dissemination of culture \cite{gonzalez2005nonequilibrium}. 

\section*{Appendix}
 \label{sec: appendix-activitydriven}
 
To find a critical threshold for the NG in activity driven networks we need to find the conditions under which solutions for $x$ and $y$ exist. The rate equations for both tell us that their solutions rely on the existence of valid solutions for $\widetilde X$, $\widetilde Y$, and $\widetilde Z$.
First we will consider our definition for $\widetilde Z = \sum_{\substack{a}}z^aa$.
To determine the existence of valid solutions for $\widetilde Z$ we need an expression for $z^a$. The definition for $z$ gives us a hint of how to define $z^a$: it is the density of agents with activity class $a$ who are not in state in $A$ or $B$:

\begin{equation}
z^a = \frac{N^a}{N} - x^a - y^a - p^a
\end{equation}
\\
The term $N^a/N$ is simply the density of all agents with activity class $a$ and with activity rates fixed \textit{a priori} this term must also be fixed. On the other hand the probability of agents being in activity class $a$ is given by the  probability distribution function $P(a)$, therefore $N^a/N = P(a)$. 
With this $z^a$ and $\widetilde Z$ become:

\begin{align}
\label{eq: za}
z^a =  & P(a) - x^a - y^a - pP(a)
\\
\widetilde Z = & \sum_{\substack{a}} z^aa = \bigl\langle a \bigr\rangle - \widetilde X - \widetilde Y - p\bigl\langle a \bigr\rangle
\end{align}

Now we see that the existence of solutions for $x$ and $y$ relies only on the existence of valid solutions for $\widetilde X$ and $\widetilde Y$. The steady states for $\widetilde X$ and $\widetilde Y$ are found through a fixed point stability analysis of their rate equations. These rate equations are obtained by multiplying (\ref{eq: xdot}) and (\ref{eq: ydot}) by $a$ and then summing over all classes to yield:

\begin{align}
\dot{\widetilde X} &= -\widetilde X\widetilde Y + \frac{1}{2}\widetilde X\widetilde Z +\frac{1}{2}\widetilde Z^2 + \widetilde Zp\bigl\langle a\bigr\rangle
\\
\dot{\widetilde Y} &= -\widetilde X\widetilde Y + \frac{1}{2}\widetilde Y\widetilde Z +\frac{1}{2}\widetilde Z^2 -\widetilde Yp\bigl\langle a\bigr\rangle
\end{align}

The steady state solutions for $\widetilde X$ and $\widetilde Y$ are then found to be:

\begin{align}
\label{eq: solXY}
\footnotesize{
(\widetilde X, \widetilde Y) =  \begin{cases}
	\quad a - ap,  0
	\\[2ex]
	\quad \dfrac{a}{6}\Bigl (1 - 7p -\sqrt{\delta}\Bigr),
	\dfrac{a}{3}\Bigl(2 - 2p + \sqrt{\delta}\Bigr)
	\\[2ex]
	\quad \dfrac{a}{6}\Bigl (1 - 7p +\sqrt{\delta}\Bigr), 
	\dfrac{a}{3}\Bigl(2 - 2p - \sqrt{\delta}\Bigr)
\end{cases}}
\\
\delta = 1 - 14p + p^2 \nonumber
\end{align}

The first of these solutions represents the trivial case where the entire population begins in the state knowing only opinion $A$ (consensus), whether they are committed or not since $\widetilde Y = 0$ and thus $y = 0$.

The other two solutions lead to stable opinion states of the population with restrictions on the value of $p$. The first of these restrictions comes from the fact that opinion states must remain physical and thus the discriminant $\delta = 1-14p+p^2 \geqslant 0$. From this two limiting values of $p$ result: $p_c = 7\pm4\sqrt{3}$, however only the lesser of these is valid as a value of $p$ since $p$ is a density and must be in the interval of $(0,1)$. 

This is the same critical threshold for $p_c$ obtained from the fixed point stability analysis of the NG on homogeneous or non-activity driven networks.

\bibliographystyle{apsrev}
\bibliography{biblio}

\end{document}